\documentclass{appolb}
\usepackage{graphicx}
\usepackage{cite}
\usepackage{bm}

\begin{document}


\title{Kaon--deuteron correlation function \\ from an effective field theory approach\thanks{Presented by Juan Torres-Rincon at the Excited QCD 2026 Workshop in Granada (Spain), on January 10, 2026.}
}

\author{Juan Torres-Rincon and \`Angels Ramos
\address{Departament de F\'isica Qu\`antica i Astrof\'isica, Universitat de Barcelona, Spain \\
Institut de Ci\`encies del Cosmos (ICCUB), Universitat de Barcelona, Spain}
}

\maketitle

\begin{abstract}
We present a study of femtoscopic correlation functions for $K^{-}d$ and $K^{+}d$ pairs, and compare our results with recent measurements by the ALICE Collaboration in both Pb-Pb and high-multiplicity $pp$ collisions. The kaon-deuteron wave functions are derived from scattering amplitudes using a unitarized chiral effective theory model describing the elementary interactions of $K^{\pm}$ mesons with nucleons. We then evaluate the $K^{\pm}d$ strong scattering amplitudes by solving the Faddeev equations within two distinct frameworks: the Impulse Approximation and the Fixed Center Approximation, which accounts for multiple scatterings. We also incorporate the long-range Coulomb effects between the kaon and the deuteron. We show that the $K^{-}d$ correlation function exhibits large sensitivity to both the size of the emitting source and the relative momentum of the pair, being heavily influenced by rescattering processes. In contrast, the $K^{+}d$ correlation function is dominated by the weakly repulsive $K^{+}N$ interaction, showing deviations from purely Coulombic behavior only at small emission source sizes. Our predictions are in agreement with the ALICE experimental data, and also with the energy-shift and width of the $1s$ level of the kaonic deuterium preliminary results from the SIDDHARTA 2 Collaboration. 
\end{abstract}

\section{Introduction}

Over the past two decades, femtoscopy has emerged as a high-precision technique for probing the space-time structure of particle-emitting sources and for detailing the final-state interactions between hadrons generated in the last stage of high-energy relativistic collisions~\cite{Heinz:1999rw, Lisa:2005dd, Fabbietti:2020bfg}. While the technique was initially established for correlations of light mesons, such as pion-pion pairs, recent experimental and theoretical focus has shifted towards more complex systems, for example, deuteron-hadron pairs. These systems can help to establish the most relevant production mechanism of light nuclei in hadronic collisions and resolve the ongoing debate between thermal freeze-out models, where nuclei emerge directly from the expanding fireball, and final-state coalescence models, where constituent nucleons merge to form nuclei at later stages~\cite{Oliinychenko:2017sfl,Mrowczynski:2019yrr}.

Correlations between a deuteron and a kaon have been successfully measured by the ALICE Collaboration, starting with the precise data on $K^{+}d$ correlations in high-multiplicity $pp$ collisions at $\sqrt{s}=13$ TeV~\cite{ALICE:2023bny}, alongside with data on both $K^{+}d$ and $K^{-}d$ correlations in Pb-Pb collisions at $\sqrt{s_{NN}}=5.02$ TeV~\cite{Rzesa:2024nra_thesis,ALICE:2026pxr}. Since a two-body treatment of kaon--deuteron systems is physically well justified at the relevant momenta~\cite{Mrowczynski:2025qys}, a robust interpretation of these complex measurements necessitates a realistic, microscopically grounded description of the kaon--deuteron interaction.

In this work, we introduce a unified, rigorous theoretical framework for kaon--deuteron femtoscopy based on chiral effective elementary interactions. We combine these amplitudes within the Faddeev equations, and compute both the $K^{-}d$ and $K^{+}d$ amplitudes. These strong amplitudes are employed, together with a separable treatment of the Coulomb interaction, to obtain the exact pair wave functions necessary for computing theoretical correlation functions to be compared directly with ALICE data~\cite{Ramos:2025ibe}.

\section{Kaon--Deuteron Scattering Amplitudes}

The determination of the $K^{-}d$ scattering amplitude, $T_{K^{-}d}$, requires summing the individual Faddeev partitions, $T_{p}^{-}$ and $T_{n}^{-}$, which contain all processes in which the incoming $K^{-}$ interacts first with the proton or the neutron of the deuteron, respectively. The system of equations reads~\cite{Kamalov:2000iy, Ramos:2025ibe},
\begin{eqnarray}
  T_{p}^{-} &=& t_{K^{-}p} + t_{K^{-}p} G_{0} T_{n}^{-} - t_{x} G_{0} T_{n}^{x} \ ,  \nonumber \\
  T_{n}^{-} &=& t_{K^{-}n} + t_{K^{-}n} G_{0} T_{p}^{-} \ , \nonumber \\
  T_{n}^{x} &=& t_{x} - t_{\overline{K}^{0}n} G_{0} T_{n}^{x} + t_{x} G_{0} T_{n}^{-} ,
  \label{eq:faddeev}
\end{eqnarray}
where $T_{n}^{x}$ accounts for the charge exchange channel $\overline{K}^0nn \leftrightarrow K^- pn$. 

In Eq.~(\ref{eq:faddeev}), the two-body elementary $s$-wave amplitudes $t_i$ refer to the elastic interactions ($t_{K^{-}p}$, $t_{K^{-}n}$, and $t_{\overline{K}^{0}n}$) and the charge exchange interaction $t_{x}$ corresponding to the transition $\overline{K}^{0}n \rightarrow K^{-}p $. These amplitudes are derived from a LO chiral unitary model~\cite{Jido:2002zk} in coupled-channels, initially developed in Ref.~\cite{Oset:1997it}. This model for the kaon--deuteron interaction has previously been employed successfully at low energy at the level of scattering lengths~\cite{Kamalov:2000iy}. In the present work, we extend the interaction to finite energy and consider the $K^{-}d$ amplitudes beyond the threshold, allowing us to account for the correlation functions at finite relative momenta, $k^*$.

The $G_{0}$ function in Eq.~(\ref{eq:faddeev}) represents the free kaon propagator convoluted by the deuteron form factor $F_d(q)$,
\begin{small}
\begin{equation}
  G_{0}(q^{0}) = \int \frac{d^{3}q}{(2\pi)^{3}} \frac{F_{d}(q)}{(q^{0})^{2} - q^{2} - m_{K}^{2} + \rm{i} \eta} \ \quad , \quad F_{d}(q) = \int d^{3}r \, e^{-\rm{i} \bm{q} \cdot \bm{r}} |\phi_{d}(r)|^{2} \ ,
\end{equation}
\end{small}
where $m_{K}$ is the intermediate kaon mass ($K^{-}$ or $\overline{K}^{0}$), and  the deuteron wave function, $\phi_{d}(r)$, contains both $s-$wave and $d$-wave contributions from the Argonne V18 NN interaction~\cite{Wiringa:1994wb}.

The system of coupled equations (\ref{eq:faddeev}) can be solved using the Impulse Approximation (IA), which amounts to exclusively keeping the first terms in $T_p^-$ and $T_n^-$; or in the Fixed Center Approximation (FCA), which results in the full solution of the algebraic system. 

In the simpler IA---or single-scattering approximation---the $K^{-}d$ amplitude, $T_{K^{-}d}^{IA}$ reads,
\begin{equation}
  T_{K^{-}d}^{IA}(k', k) = [t_{K^{-}p}(k', k) + t_{K^{-}n}(k', k)] F_{d}(q),
\end{equation}
where $q = |\bm{k} - \bm{k}'|/2$ is half the momentum transferred to the deuteron, defined from the initial kaon momentum $\bm{k}$ and final momentum $\bm{k}'$. 

The full FCA equations yield a more complex amplitude containing all orders of rescattering,
\begin{small}
\begin{equation}
  T_{K^{-}d}^{FCA} = \frac{T_{K^{-}d}^{IA} + (2t_{K^{-}p}t_{K^{-}n} - \tilde{t}_{x}^{2})G_{0} - 2\tilde{t}_{x}^{2} t_{K^{-}n} G_{0}^{2}}
  {1 - t_{K^{-}p}t_{K^{-}n} G_{0}^{2} + \tilde{t}_{x}^{2} t_{K^{-}n} G_{0}^{3}} , \quad \textrm{with} \quad \tilde{t}_{x} \equiv \frac{t_{x}}{\sqrt{1 + t_{\overline{K}^{0}n}G_{0}}} \ .
\end{equation}
\end{small}

By comparing the FCA predictions with full exact Faddeev calculations for near-threshold observables~\cite{Ramos:2025ibe}, we can confidently assign a theoretical uncertainty to the FCA on the order of $10\%-30\%$. This uncertainty lies well within the systematic uncertainties connected to the use of the chiral model used in the elementary two-body matrix elements.

For the $K^{+}d$ scattering amplitude, one can obtain equivalent mathematical expressions by replacing $K^{-} \rightarrow K^{+}$, $\overline{K}^{0} \rightarrow K^{0}$, $p \rightarrow n$, and $n \rightarrow p$. 

The energy dependence of the $K^-d$ scattering amplitude~\cite{Ramos:2025ibe} reveals a pronounced peak just below the $K^{-}pn$ threshold. This structure is a reflection of the subthreshold $\Lambda(1405)$ resonance, dynamically generated in the $I=0$ $\bar{K}N$ interaction. The FCA amplifies the strength of this state with respect to the IA due to the multiple rescattering within the deuteron. Conversely, the $K^{+}d$ amplitude is essentially featureless and inherently weak in strength (both in the IA and the FCA) aligning with the expectation that the $K^{+}N$ interaction is mildly repulsive and elastic at low energies.

\begin{table}[ht]
\centering
\renewcommand{\arraystretch}{1.3}
\begin{tabular}{lcccc}
\hline\hline
 & $A_{K^{-}d}$ (fm) & $A_{K^{+}d}$ (fm) & $\epsilon_{1s}$ (eV) &  $\Gamma_{1s}$ (eV)\\
\hline
IA & $-0.586 + \textrm{i}2.145$ & -0.47 & 791 & 2063 \\
\hline
FCA & $-2.062 + \textrm{i}1.767$ & -0.43 & 1124 & 626 \\
\hline\hline
\end{tabular}
\caption{Calculated results for the $K^{-}d$ and $K^{+}d$ scattering lengths, alongside predictions for the strong-interaction energy shift ($\epsilon_{1s}$) and width ($\Gamma_{1s}$) of the $1s$ atomic level in kaonic deuterium.}
\label{tab:scattering_lengths}
\end{table}

In Table~\ref{tab:scattering_lengths} we present results of the scattering lengths at the $K^\pm d$ thresholds, for both IA and FCA approximations. The large real negative part of the $K^{-}d$ scattering length is a signature of the  dynamically generated $\bar{K}NN$ system. The large imaginary part reflects the available phase space for conversion channels ($\pi \Lambda N$, $\pi \Sigma N$). In the case of the $K^{+}d$ scattering lengths, the predictions from the IA and FCA frameworks are practically identical due to its weaker interaction.

In the same table we present our predictions for the energy shift ($\epsilon_{1s}$) and total width ($\Gamma_{1s}$) of the $1s$ orbital level in exotic kaonic deuterium atom. These quantities present large differences between IA and FCA. In the latter case, the decay width $\Gamma_{1s}$ is in striking agreement with preliminary estimates from the SIDDHARTA-2 experimental collaboration~\cite{SIDDHARTA-2_Sgaramella,Artibani:2026kvw}, though the calculated energy shift is approximately $30\%$ larger. 

\section{Correlation Functions}

The application of the two-body amplitude to the femtoscopic correlation function between a $K^{\pm}$ and a $d$ uses the Koonin-Pratt formula~\cite{Koonin:1977fh,Pratt:1990zq},
\begin{equation}
  C(k^*) = \int {\rm d}^3r \ S_{12}(r) |\Psi(r; k^*)|^2 \ ,
  \label{eq:koonin_pratt}
\end{equation}
where $\Psi(r; k^*)$ represents the relative two-body wave function and $S_{12}(r)$ is the normalized relative source function, which follows a Gaussian profile,
\begin{equation}
  S_{12}(r) = \frac{1}{(2\pi R_{Kd}^2)^{3/2}} \exp{ \left( -\frac{r^2}{2R_{Kd}^2} \right)} \ ,  \label{eq:source}
\end{equation}
where the two-body source size $R_{Kd}$ is related to the individual $K$ and $d$ source radii. Here, it is extracted from the experimental analysis of~\cite{Rzesa:2024nra_thesis}.

We restrict the low-energy strong dynamics to the $L=0$ partial wave, but to all orders in the Coulomb interaction. In this case, we separate the full wave function into the complete Coulomb wave function, its $s-$wave projection, and the Coulomb + strong $s$-wave component: $\Psi(r; k^*) = \Phi^{\rm{c}}(r; k^*) - \Phi_{0}^{\rm{c}}(k^*r) + \Psi_{0}(r; k^*)$. The correlation function reads,
\begin{small}
\begin{equation}
  C(k^*) = \int {\rm d}^3r S_{12}(r) |\Phi^{\rm{c}}(r; k^*)|^2 + \int 4\pi r^2 {\rm d}r S_{12}(r) \left( |\Psi_{0}(r; k^*)|^2 - |\Phi_{0}^{\rm{c}}(k^*r)|^2 \right) \ .
\end{equation}
\end{small}

The strong + Coulomb wave function $\Psi_{0}(r; k^*)$ is determined from the Lippmann-Schwinger equation projected into the $s$-wave:
\begin{equation}
  \Psi_{0}(r; k^*) = j_0(k^* r) + \int_{0}^{\infty} \frac{{\rm d}^3q}{(2\pi)^3} \frac{M_d}{E_d} \frac{1}{2\omega_K} \frac{T_{Kd}(q, k^*; \sqrt{s_{Kd}}) \ j_0(qr)}{\sqrt{s_{Kd}} - E_d - \omega_K + \rm{i}\eta} \ ,
\end{equation}
where $j_0(x)$ is the spherical Bessel function, $\omega_K = \sqrt{q^2 + m_K^2}$ represents the intermediate kaon energy, $E_d$ is the corresponding deuteron energy, and $\sqrt{s_{Kd}}$ defines the total invariant energy of the pair in the CoM frame.

To simplify the calculation the strong and Coulomb scattering amplitudes are assumed to be approximately separable, $T_{Kd} \approx T_{Kd}^{FCA} + T_{Kd}^{\rm{c}}$ (estimated within a 10 \% error by comparing the results with the solution of the Schr\"odinger equation of the same problem). In addition, we apply the on-shell factorization on the strong scattering amplitude within the momentum integral, which we regularize with a momentum cutoff $\Lambda = 1000\textrm{ MeV}$.

\begin{figure}[htb]
\centerline{%
\includegraphics[width=12.5cm]{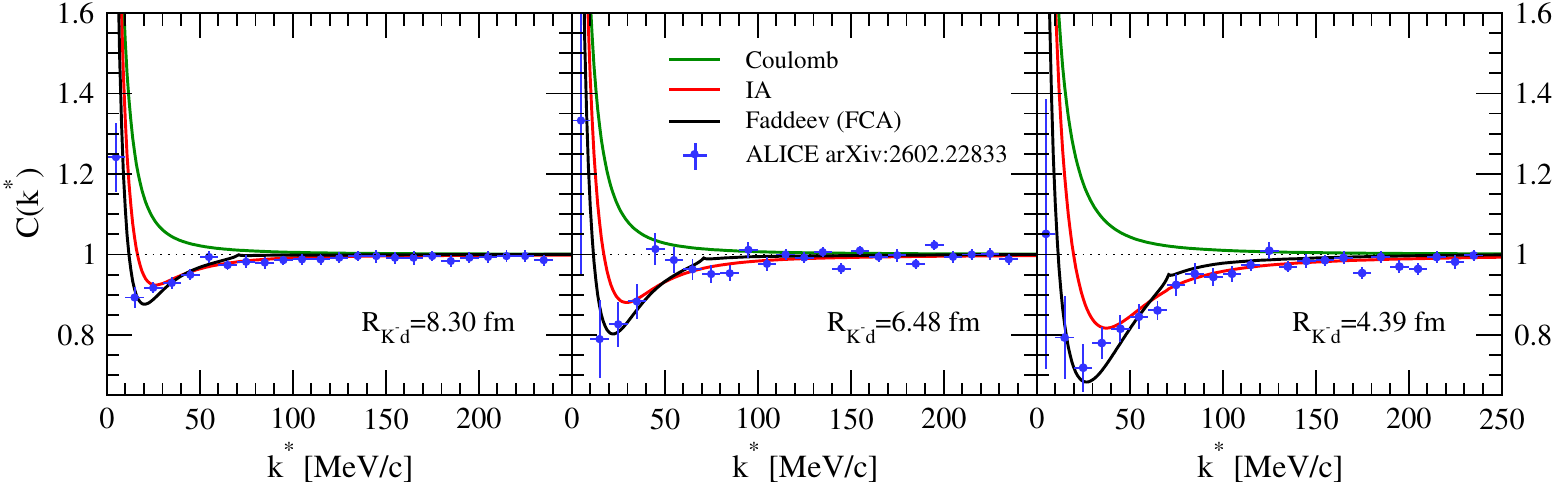}}
\centerline{
\includegraphics[width=12.5cm]{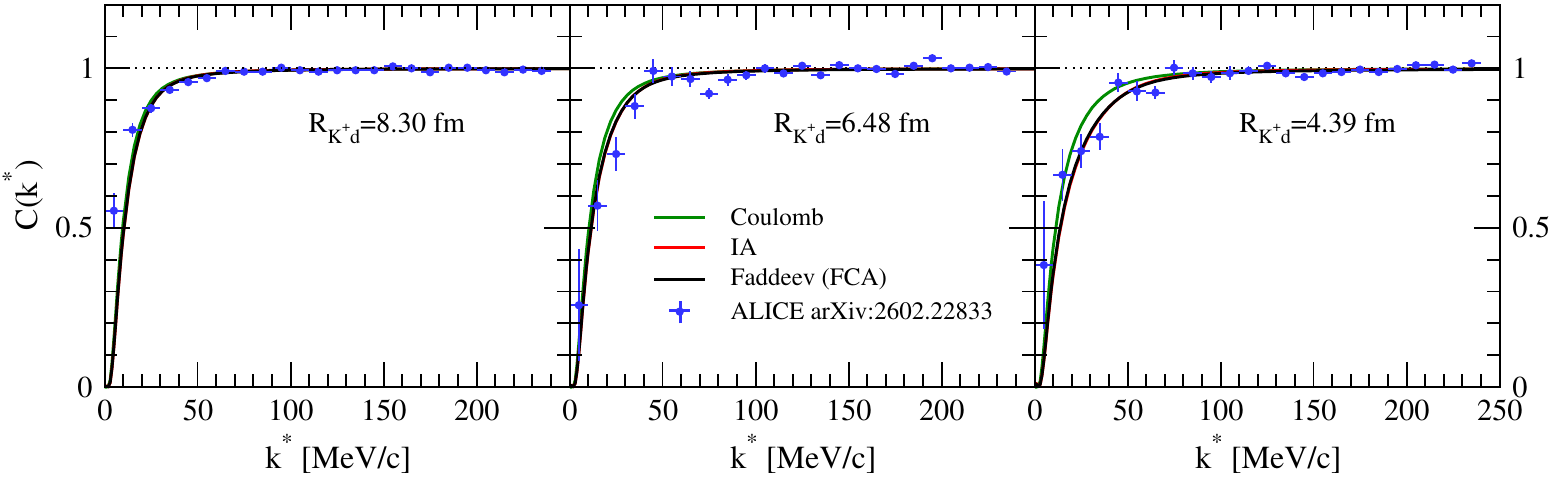}}
\caption{Top: $K^-d$ correlation functions as obtained from our models (IA and FCA) and from ALICE data in Pb+Pb collisions at $\sqrt{s_{NN}} =5.02$ TeV at different centralities (0-10 \%, 10-30 \%, 30-50\%). Bottom: Same for $K^+d$ correlation functions.}
\label{fig:corre}
\end{figure}

\begin{figure}[htb]
\centerline{%
\includegraphics[width=6cm]{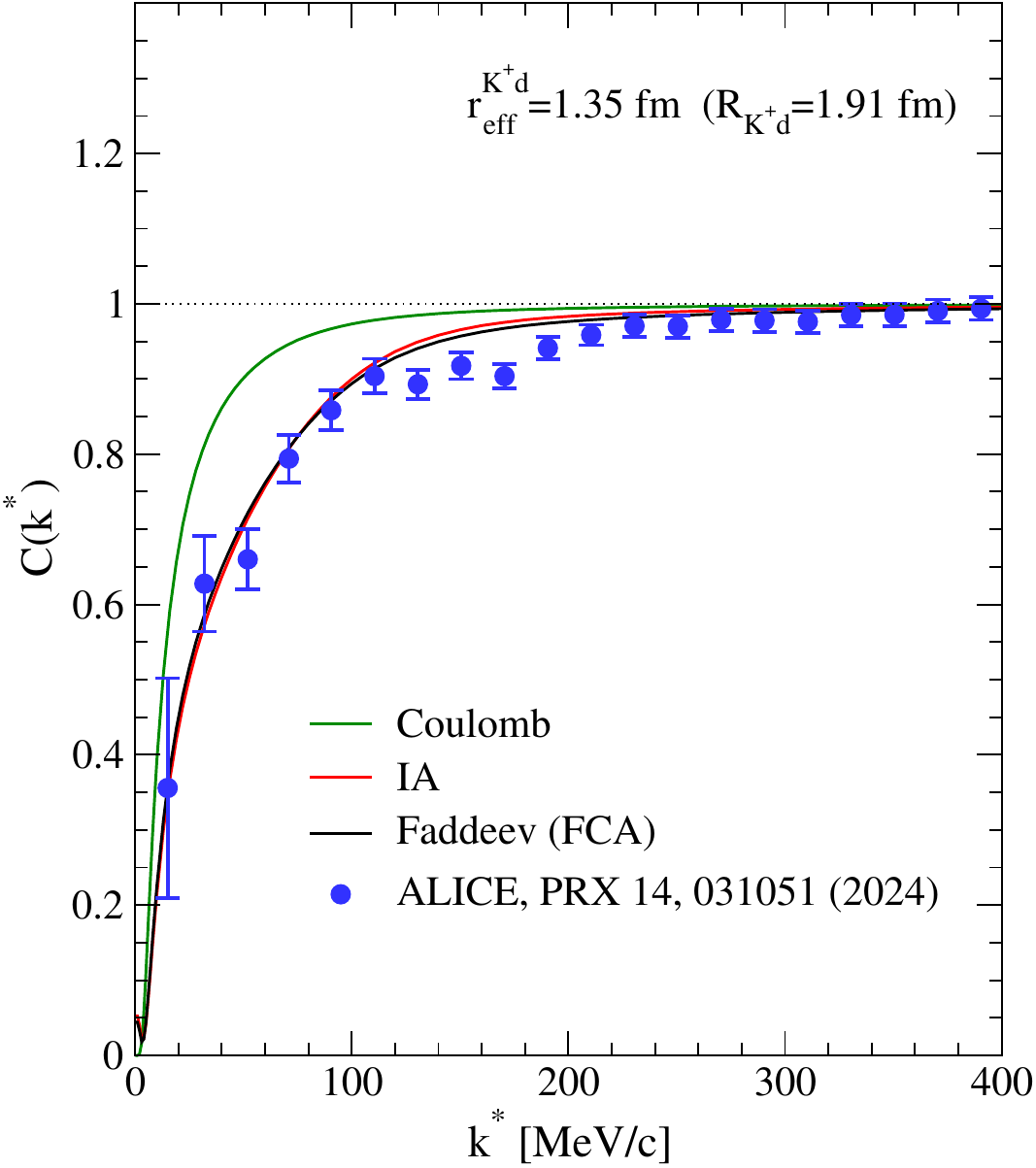}}
\caption{$K^+d$ correlation function of obtained from our models (IA and FCA) and from ALICE data in $pp$ collisions at $\sqrt{s} =13$ TeV.}
\label{fig:corre2}
\end{figure}

In Fig.~\ref{fig:corre} we contrast our theoretical calculations for the $K^{-}d$ (top) and $K^+d$ (bottom) correlation functions against ALICE Pb-Pb collision data ($\sqrt{s_{NN}}=5.02$ TeV) across three centrality classes ($0-10\%$, $10-30\%$, and $30-50\%$ corresponding to distinct emitting source radii $R_{Kd} \approx 8.30\textrm{ fm}, 6.48\textrm{ fm},$ and $4.39\textrm{ fm}$). In the $K^-d$ case, the Coulomb interaction alone is unable to describe the data. When the strong interaction is included, the correlation function presents a depletion, which is inherited from the subthreshold generated state. The FCA significantly improves over the simplistic IA, yielding a better explanation of the ALICE data. For the $K^{+}d$ correlation functions, the strong interaction correction is tiny with respect to Coulomb, and essentially any scenario can describe the ALICE data for the three centralities. In this system, much smaller sources---high-multiplicity $pp$ collisions at $\sqrt{s}=13$ TeV---reveal that the addition of the weakly repulsive $K^{+}d$ strong interaction is required to match the high-precision experimental measurements. These can be clearly seen in Fig.~\ref{fig:corre2}. In the figure we show the Gaussian radius $r_{{\rm eff}}^{K^+ d} \equiv R_{K^+d}/\sqrt{2}$, which is the one quoted in Ref.~\cite{ALICE:2023bny}. It is related by a factor $\sqrt{2}$ to ours since they use a Gaussian source $S_{12}(r) \propto \exp(-r^2/(4 r_{Kd}^2))$, instead of our Eq.~(\ref{eq:source}).

\section{Conclusions}

We have developed a unified framework designed to analyze $K^{-}d$ and $K^{+}d$ femtoscopy. Our approach combines realistic chiral effective field theory interactions in the elementary $K^{-}N$ and $K^{+}N$ sectors with Faddeev equations, evaluated under both the impulse approximation and the fixed-center approximation, to account for the kaon--deuteron interaction. The model includes both strong and Coulomb interactions, combined into the Koonin-Pratt formalism to obtain theoretical two-particle femtoscopic correlation functions.

The $K^{-}d$ system is dominated by a strong attraction able to generate a subthreshold resonance, which is the three-body manifestation of the $\Lambda(1405)$ states. In this system, multiple-scattering effects are dominant and can only be captured by the FCA, which provides a good agreement with femtoscopy ALICE data in heavy-ion collisions at all centralities. Conversely, the $K^{+}d$ system is governed by a weakly repulsive, predominantly elastic potential. Therefore, there are essentially no differences between the IA and the FCA. In fact, departures from the pure Coulomb interaction are not visible in heavy-ion collision data at any centrality. Only in small systems---high-multiplicity $pp$ collisions---a deviation from Coulomb, favouring the inclusion of the strong interaction, is seen in the ALICE data.

Finally, predictions concerning the $1s$ atomic level energy shift and decay width of exotic kaonic deuterium corroborate the importance of multiple scatterings, bringing our theoretical prediction into closer alignment with preliminary empirical data from the SIDDHARTA-2 experiment.

Future theoretical refinements will focus on improving the unitarity of the three-body system by accounting from coherent propagation of the kaon--deuteron systems~\cite{Agatao:2025ckp}, and refining the factorization assumption between the strong and Coulomb amplitudes.

\section*{Acknowledgments}
This work has been supported by the projects CEX2024-001451-M (Unidad de Excelencia "María de Maeztu") and PID2023-147112NB-C21, funded by the Spanish MCIN/ AEI/10.13039/501100011033/, and by Contract 2021 SGR 171 by the Generalitat de Catalunya. J. M. T.-R. also acknowledges support from Grant No. 402942/2024-8 by the Brazilian CNPq (National Council for Scientific and Technological Development).

\bibliographystyle{unsrt}
\bibliography{references}

\begin{thebibliography}{10}

\bibitem{Heinz:1999rw}
Ulrich~W. Heinz and Barbara~V. Jacak.
\newblock {Two particle correlations in relativistic heavy ion collisions}.
\newblock {\em Ann. Rev. Nucl. Part. Sci.}, 49:529--579, 1999.

\bibitem{Lisa:2005dd}
Michael~Annan Lisa, Scott Pratt, Ron Soltz, and Urs Wiedemann.
\newblock {Femtoscopy in relativistic heavy ion collisions}.
\newblock {\em Ann. Rev. Nucl. Part. Sci.}, 55:357--402, 2005.

\bibitem{Fabbietti:2020bfg}
L.~Fabbietti, V.~Mantovani~Sarti, and O.~Vazquez~Doce.
\newblock Study of the strong interaction among hadrons with correlations at
  the lhc.
\newblock {\em Ann. Rev. Nucl. Part. Sci.}, 71:377--402, 2021.

\bibitem{Oliinychenko:2017sfl}
Dmytro Oliinychenko.
\newblock {\em {Interfaces between relativistic hydrodynamics and transport for
  the dynamical description of heavy ion collisions}}.
\newblock PhD thesis, Frankfurt U., 7 2017.

\bibitem{Mrowczynski:2019yrr}
Stanislaw Mr\'owczy\'nski and Patrycja S\l{}o\'n.
\newblock {Hadron\textendash{}Deuteron Correlations and Production of Light
  Nuclei in Relativistic Heavy-ion Collisions}.
\newblock {\em Acta Phys. Polon. B}, 51(8):1739--1755, 2020.

\bibitem{ALICE:2023bny}
Shreyasi Acharya et~al.
\newblock {Exploring the Strong Interaction of Three-Body Systems at the LHC}.
\newblock {\em Phys. Rev. X}, 14(3):031051, 2024.

\bibitem{Rzesa:2024nra_thesis}
Wioleta Rzesa.
\newblock {\em Non-identical particle femtoscopy of pairs containing deuteron
  and interaction studies of nucleons with strange matter}.
\newblock {Ph.D.} thesis, Warsaw U. of Tech., 2024.

\bibitem{ALICE:2026pxr}
Dana Ali Hassan~Abdallah et~al.
\newblock {First measurement of the strong interaction scattering parameters
  for the $\mathbf{K^-d}$ and $\mathbf{K^+d}$ systems}.
\newblock 2026.

\bibitem{Mrowczynski:2025qys}
Stanislaw Mrowczynski.
\newblock {Two- versus three-body approach to femtoscopic hadron-deuteron
  correlations}.
\newblock {\em Phys. Lett. B}, 864:139413, 2025.

\bibitem{Ramos:2025ibe}
{\`A}ngels Ramos, Juan~M. Torres-Rincon, Alejandro de~Fagoaga, and Esteve
  Cabr{\'e}.
\newblock {Kaon-deuteron femtoscopy from unitarized chiral interactions}.
\newblock {\em Phys. Rev. D}, 113(3):036020, 2026.

\bibitem{Kamalov:2000iy}
S.~S. Kamalov, E.~Oset, and A.~Ramos.
\newblock {Chiral unitary approach to the K- deuteron scattering length}.
\newblock {\em Nucl. Phys. A}, 690:494--508, 2001.

\bibitem{Jido:2002zk}
D.~Jido, E.~Oset, and A.~Ramos.
\newblock {Chiral dynamics of p wave in K- p and coupled states}.
\newblock {\em Phys. Rev. C}, 66:055203, 2002.

\bibitem{Oset:1997it}
E.~Oset and A.~Ramos.
\newblock {Nonperturbative chiral approach to s wave anti-K N interactions}.
\newblock {\em Nucl. Phys. A}, 635:99--120, 1998.

\bibitem{Wiringa:1994wb}
Robert~B. Wiringa, V.~G.~J. Stoks, and R.~Schiavilla.
\newblock {An Accurate nucleon-nucleon potential with charge independence
  breaking}.
\newblock {\em Phys. Rev. C}, 51:38--51, 1995.

\bibitem{SIDDHARTA-2_Sgaramella}
Francesco Sgaramella et~al.
\newblock {\it Kaonic Atoms X-ray Spectroscopy with SIDDHARTA-2: The First
  Measurement of Kaonic Deuterium}, talk presented at the Workshop on
  Fundamental Physics with Exotic Atoms, INFN Frascati, Italy, 2025.

\bibitem{Artibani:2026kvw}
Francesco Artibani et~al.
\newblock {Kaonic atoms measurements with the SIDDHARTA-2 experiment at
  DA$\Phi$NE}.
\newblock {\em PoS}, HADRON2025:172, 2026.

\bibitem{Koonin:1977fh}
S.~E. Koonin.
\newblock {Proton Pictures of High-Energy Nuclear Collisions}.
\newblock {\em Phys. Lett. B}, 70:43--47, 1977.

\bibitem{Pratt:1990zq}
S.~Pratt, T.~Csorgo, and J.~Zimanyi.
\newblock {Detailed predictions for two pion correlations in ultrarelativistic
  heavy ion collisions}.
\newblock {\em Phys. Rev. C}, 42:2646--2652, 1990.

\bibitem{Agatao:2025ckp}
Breno Agat{\~a}o, Pedro Brand{\~a}o, A.~Mart{\'\i}nez~Torres, K.~P.
  Khemchandani, Luciano~M. Abreu, and E.~Oset.
\newblock {Correlation functions for $n\,\bar{D}_{s1}(2460)$ and
  $n\,\bar{D}_{s1}(2536)$}.
\newblock {\em Eur. Phys. J. C}, 85(10):1136, 2025.

\end{thebibliography}

\end{document}